# How Memristors Work: A Step-by-Step Guide to the CALM Model

*Support text for a self-study tutorial: introduction, instructions, summary, references and solutions to the exercises*

Tutorial materials and Python code: https://doi.org/10.5281/zenodo.23463175

## Juan Bisquert

Instituto de Tecnología Química (ITQ), Consejo Superior de Investigaciones Científicas-Universitat Politècnica de València, 46022, Valencia, Spain

Email: jbisquer@itq.upv.es

## Summary

Memristors made of very different materials, halide perovskites, metal oxides and nanofluidic channels, show a common set of dynamical features: two limiting conductances, hysteresis that depends on the scan rate, regular and inverted hysteresis loops, large and negative capacitances and inductive loops in impedance spectra, and potentiation and depression under pulse trains. This tutorial shows how a single minimal model, the conductance-activated quasi-linear memristor (CALM) model, explains all these features with three ingredients: the two conductance states, the rectification function $x_{eq}(u)$ and the characteristic time $\tau(u)$. The tutorial is a slide presentation that the reader can follow alone, with exercises designed to be solved with the help of an AI assistant, reference Python code, and this support text. The text can also be read on its own: Section 4 gives a compact account of the content of the slides, with the main equations and figures, and the solutions to the exercises are given at the end.

## 1. Introduction

Memristors are usually introduced through their applications, as binary memories with a high resistance state (HRS) and a low resistance state (LRS), or as artificial synapses whose conductance changes gradually with the applied pulses. Most researchers who work with these devices are familiar with the phenomenology: the SET and RESET transitions, volatile and non-volatile operation, bipolar and unipolar switching. The difficulty appears when the measured curves must be interpreted quantitatively. Why does the hysteresis loop change with the scan rate? Why are some loops inverted? Why does the impedance spectrum show a capacitance thousands of times larger than the geometric capacitance, or an inductive loop? Why does the first cycle differ from the following ones? Why is a measured loop not pinched at the origin, as the textbooks say it should be?

All these questions have a common answer. A memristor has a fast variable, the voltage $u$, and at least one slow internal variable $x$ that controls the conductance [1]. Every dynamical effect discussed in this tutorial comes from the fact that the internal state lags behind the voltage. The CALM model is the simplest model that gives this lag a concrete form and still reproduces the experiments. It was developed for nanofluidic pores [2] and for volatile and non-volatile halide perovskite memristors [3], and has been applied to lead-free bismuth halide memristors [4] and multipore nanofluidic membranes [5].

The audience of the tutorial is formed by researchers and doctoral students who already know memristors and want to understand the characteristics of a model and apply it to their own data. No previous knowledge of nonlinear dynamics is required; familiarity with impedance spectroscopy helps in Part 6 but is not essential.

## 2. How to use this tutorial

### 2.1 Materials

The tutorial consists of three elements.

(i) The slide presentation *CALM memristor tutorial v3.pptx* (65 slides). It is designed to be followed without a lecturer. Each slide develops one idea: a short explanatory text, one figure, and a key idea that summarizes the slide. The model is built step by step, starting from the physical picture of a fast voltage and a slow internal state. A few slides from published work show the experimental evidence, with the source given on each of them.

(ii) This support text, with an introduction, the structure of the course, a written summary of the content, the list of references and the solutions to the exercises.

(iii) The reference code *calm_tutorial.py*, which produced every simulated figure of the slides and of this text. It requires only Python with NumPy, SciPy and Matplotlib, and it integrates the state equations with an exponential update that is stable even for very abrupt switching.

### 2.2 Exercises with an AI assistant

The exercises are marked with the badge *ASK YOUR AI*. Each exercise gives a prompt that the reader pastes into an AI assistant, preferably one that can execute Python code, so that the model responds visibly to the changes of its parameters. The purpose is not to obtain an answer but to see the dynamics: before running each variation, predict the result, then compare. Each exercise slide lists *what to look for*, and the solutions are collected at the end of this text (Sections S1–S9). If the AI produces a result that differs from the solution, compare its code with *calm_tutorial.py*; a frequent source of error is an explicit Euler integration with a time step longer than the shortest relaxation time.

### 2.3 Suggested pace

Parts 1–3 build the model and explain the hysteresis; they should be read in order and take about half a day with the exercises. Parts 4–8 apply the same ideas to device classes, impedance, pulses and more complex devices, and can be read in any order. Readers who prefer text can read Section 4 first and use the slides as an atlas of figures.

## 3. Structure of the tutorial

Table 1 lists the eight parts of the presentation, with the corresponding slides and exercises.

**Table 1. Structure of the tutorial.**

| Part | Content | Slides | Exercises |
|---|---|---|---|
| 0 | Title, about this tutorial, plan | 1–3 | |
| 1 | The idea: a fast voltage and a slow internal state | 4–8 | |
| 2 | Building the model step by step: two conductances, target state, stationary curve, relaxation time; tribute to Hodgkin and Huxley | 9–19 | 1, 2 |
| 3 | Hysteresis: capacitive and inductive types, the lag of the state, scan rate, cycling | 20–30 | 3, 4 |
| 4 | One model, many devices: volatility, abrupt and gradual, bipolar and unipolar | 31–35 | 5 |
| 5 | Small signals: transients, conduction inductance and capacitance, impedance | 36–43 | 6 |
| 6 | Pulses: potentiation, depression and frequency-dependent plasticity | 44–49 | 7 |
| 7 | Beyond the simplest picture: non-pinched loops, two processes, crossings, dual-inductive devices | 50–58 | 8, 9 |
| 8 | Using the model with your data: workflow, pitfalls, summary, references, collaborators | 59–65 | |

## 4. Summary of the content

### 4.1 What experiments show

Memristors have been made with many materials, for example halide perovskites [6,7], $Al_2O_3$ resistive memories [8] and nanofluidic conical pores [5,9]. Despite the different physical mechanisms, filament formation, valence change, interfacial ion accumulation or ion concentration polarization in a nanochannel, the curves share the features listed in the Summary. A particularly instructive example is the volatile $MAPbBr_3$ memristor of ref [10], whose impedance changes from two capacitive arcs at low bias to an inductive loop at forward bias, while the current rises slowly after a voltage step. A model should explain all these observations with a few parameters. The usual classification of memristors by polarity, volatility and switching dynamics [11] will appear naturally in Section 4.4 as different values of the model parameters.

### 4.2 The CALM model

The CALM model describes the current $I$ through a device at voltage $u$ by [2,3]

$$I = [g_L + (g_H - g_L)\,x]u + C_m \frac{du}{dt} \tag{1}$$

$$\tau(u)\frac{dx}{dt} = x_{eq}(u) - x \tag{2}$$

The internal state $x$ is normalized between 0 and 1. The conductance $G(x) = g_L + (g_H - g_L)x$ interpolates linearly between the low conductance $g_L$ and the high conductance $g_H$, and $C_m$ is the geometric or membrane capacitance in parallel. The model is quasi-linear because the equations are linear in $x$; the nonlinearity resides in the voltage dependence of the equilibrium state $x_{eq}(u)$ and of the relaxation time $\tau(u)$. The equilibrium state is a sigmoid

$$x_{eq}(u) = \frac{1}{1+\exp[-(u-V_B)/V_m]} \tag{3}$$

with threshold $V_B$ and width $V_m$. We call it the rectification function, because it selects which conductance applies at each voltage: even without any dynamics, the stationary current $I_{ss}(u) = G\left(x_{eq}(u)\right) u$ moves from the line $g_L u$ to the line $g_H u$ across the threshold (Fig. 1a). A small $V_m$ gives an abrupt transition and a large one a gradual transition (Fig. 1b). For a unipolar device the argument of the sigmoid is $|u|$.

The relaxation time has a bell or window shape [3],

$$\tau(u) = \tau_{min} + (\tau_{max} - \tau_{min})\, W(u) \tag{4}$$

with the window function

$$W(u) = s\left(\frac{u-V_{OFF}}{w_{OFF}}\right) - s\left(\frac{u-V_{ON}}{w_{ON}}\right), \qquad s(z) = \frac{1}{1+e^{-z}} \tag{5}$$

The widths $w_{OFF}$ and $w_{ON}$ set the sharpness of the two edges (they are called $V_-$ and $V_+$ in ref [3]). Inside the window $V_{OFF} < u < V_{ON}$ the time is long, $\tau_{max}$, and outside it is short, $\tau_{min}$. Where $\tau$ is short the dynamics is activated and $x$ follows $x_{eq}$ quickly. If the window contains $u = 0$, the state is frozen at zero bias and the device is non-volatile; if $\tau(0)$ is short, the device forgets and is volatile (Fig. 1c).

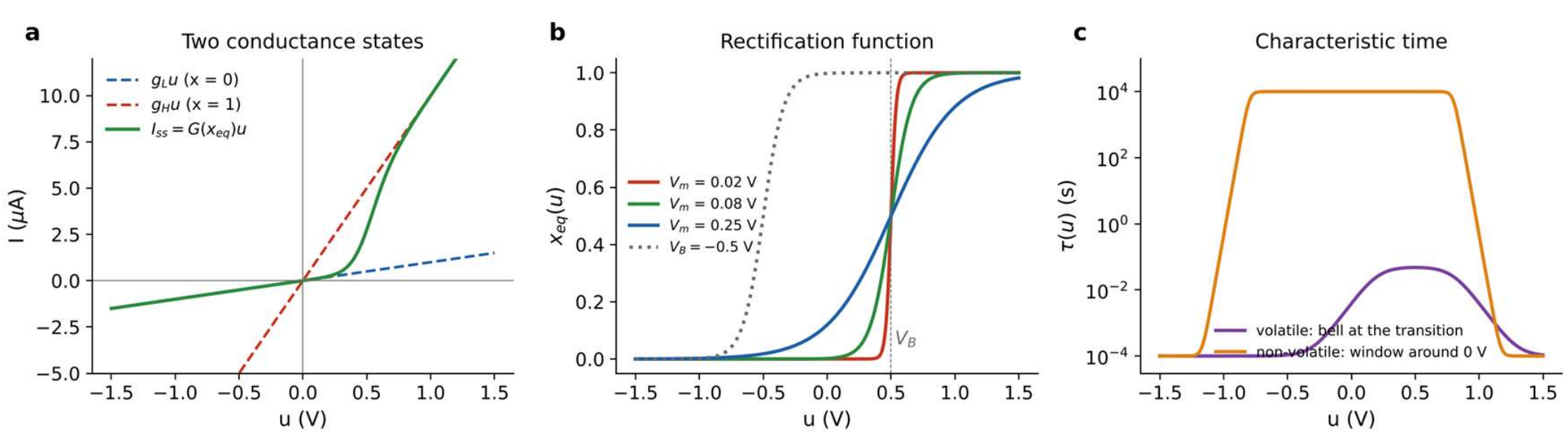

**Figure 1.** The three ingredients of the CALM model. (a) Lines $g_L u$ and $g_H u$ and stationary current $I_{ss}$ for $g_L$ = 1 µS, $g_H$ = 10 µS, $V_B$ = 0.5 V and $V_m$ = 0.08 V. (b) Rectification function for three values of $V_m$, and for $V_B$ = −0.5 V. (c) Relaxation time of a volatile device (bell centred on the transition) and of a non-volatile device (long window around 0 V).

The structure of the CALM model is not new. In 1952 Hodgkin and Huxley described the action potential of the squid giant axon with ionic conductances controlled by gating variables $m$, $h$ and $n$, each of which relaxes toward a voltage-dependent steady state with a voltage-dependent time constant [12]. Their steady states are sigmoids and their time constants are bell-shaped functions of the voltage, which are the rectification function and the relaxation time of the CALM model. The delayed activation of the potassium conductance is a conduction inductor, and an inductive impedance of the squid axon had been measured by Cole and Baker in 1941 [13]. Chua and co-workers later showed that the potassium and sodium channels of the Hodgkin–Huxley model are memristors [14]. In this sense, the Hodgkin–Huxley equations are the first memristor model, and the CALM model is a minimal member of the same family.

When the voltage is swept very slowly, $x = x_{eq}(u)$ at every voltage and the current follows $I_{ss}(u)$ without hysteresis (green curves in Fig. 2a and 2b). At faster sweeps the state lags behind $x_{eq}$ and the forward and reverse branches separate (Fig. 2b). Hysteresis is the trace of the lag of the internal state.

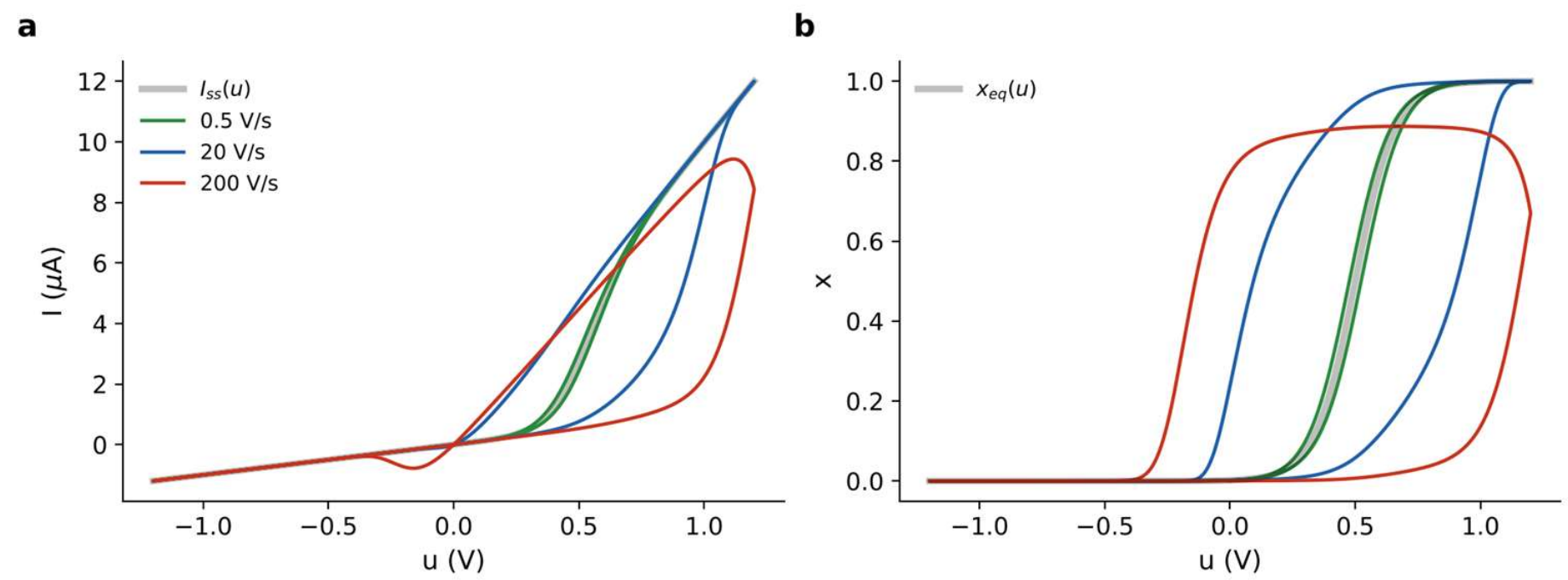

**Figure 2.** Quasi-static limit and origin of the hysteresis in a volatile CALM device. (a) Current–voltage curves at three scan rates compared with $I_{ss}$. (b) State $x$ along the sweep compared with $x_{eq}(u)$.

### 4.3 Hysteresis, scan rate and cycling

Hysteresis appears when the current is measured during a voltage sweep and some internal process is slower than the sweep. If the device responded instantly, the forward and reverse branches would both coincide with the stationary current, the current obtained after waiting at each voltage. With a slow internal process the current depends on the sweep rate and on the direction of the sweep. Two basic types of hysteresis can be distinguished [15]. In capacitive hysteresis the forward current is above the stationary curve and the reverse current below it, so that the loop turns clockwise, and the impedance spectrum shows a positive arc. In inductive hysteresis the forward current is below the stationary curve and the loop turns counterclockwise, and the impedance shows a negative arc, an inductive loop. A real capacitor connected in parallel with a resistor produces capacitive hysteresis, because the charging current adds to the current on the way up and subtracts on the way down (Fig. 3a). A real inductor in a parallel branch produces inductive hysteresis, because the current in the coil lags behind the voltage (Fig. 3b). A memristor shows both types in different voltage ranges, although it contains neither a capacitor nor an inductor (Fig. 3c). In this case both originate in the same process, the delayed change of the conduction, and are called conduction capacitance and conduction inductance [16].

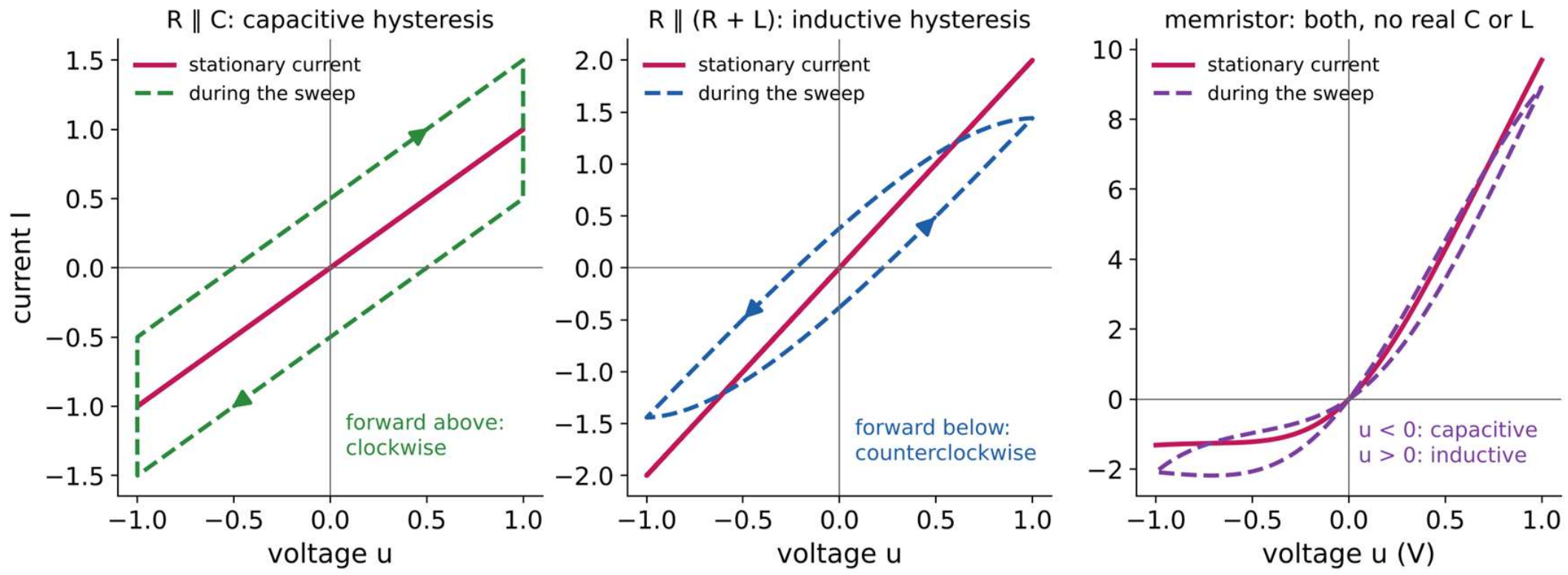

**Figure 3.** Origin of capacitive and inductive hysteresis. (a) Resistor in parallel with a capacitor. (b) Resistor in parallel with a resistor–inductor branch. (c) CALM memristor, capacitive at negative and inductive at positive voltage. The solid line is the stationary current and the dashed line the current during a triangular sweep.

The hysteresis loop vanishes at both limits of the scan rate (Fig. 4a). At very slow sweeps $x$ follows $x_{eq}$; at very fast sweeps $x$ has no time to move and the current follows a straight line with the initial conductance. The loop area is maximum at an intermediate rate, when the time spent crossing the transition region is comparable with $\tau$ (Fig. 4b). The position of the maximum is therefore a measure of the relaxation time, which must agree with the time obtained from impedance spectroscopy (Section 4.5). The experimental observation that fast sweeps outpace the internal relaxation and collapse the loop [8] is the right half of Fig. 4b.

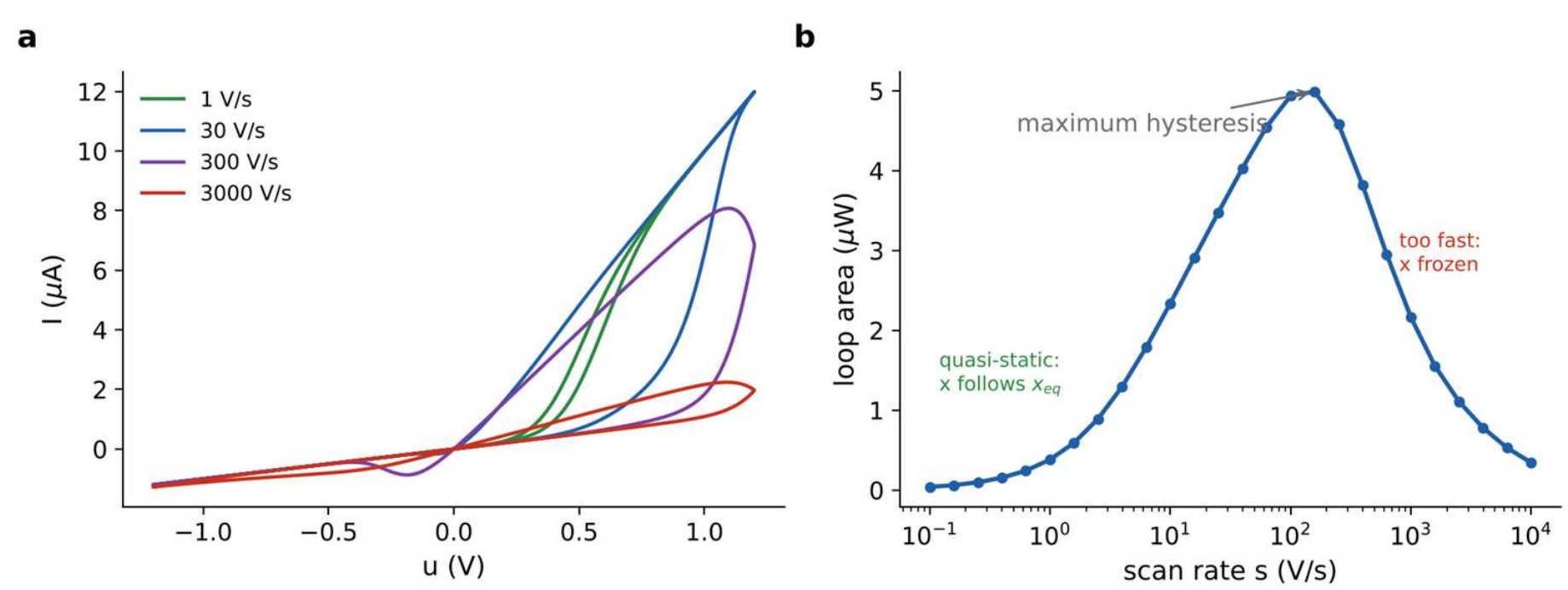

**Figure 4.** Scan-rate dependence of the CALM hysteresis. (a) Loops at four scan rates. (b) Loop area, sum of the absolute areas of the lobes, as a function of the scan rate.

In a non-volatile device the state is not reset between cycles, and the loop drifts from cycle to cycle (Fig. 5a) until the change of $x$ during SET is compensated by the change during RESET (Fig. 5b). The first cycle is a transient and the stabilized loop is a limit cycle. Reported data should state which cycle is shown, and the model should be fitted to the stabilized cycle or to the full history.

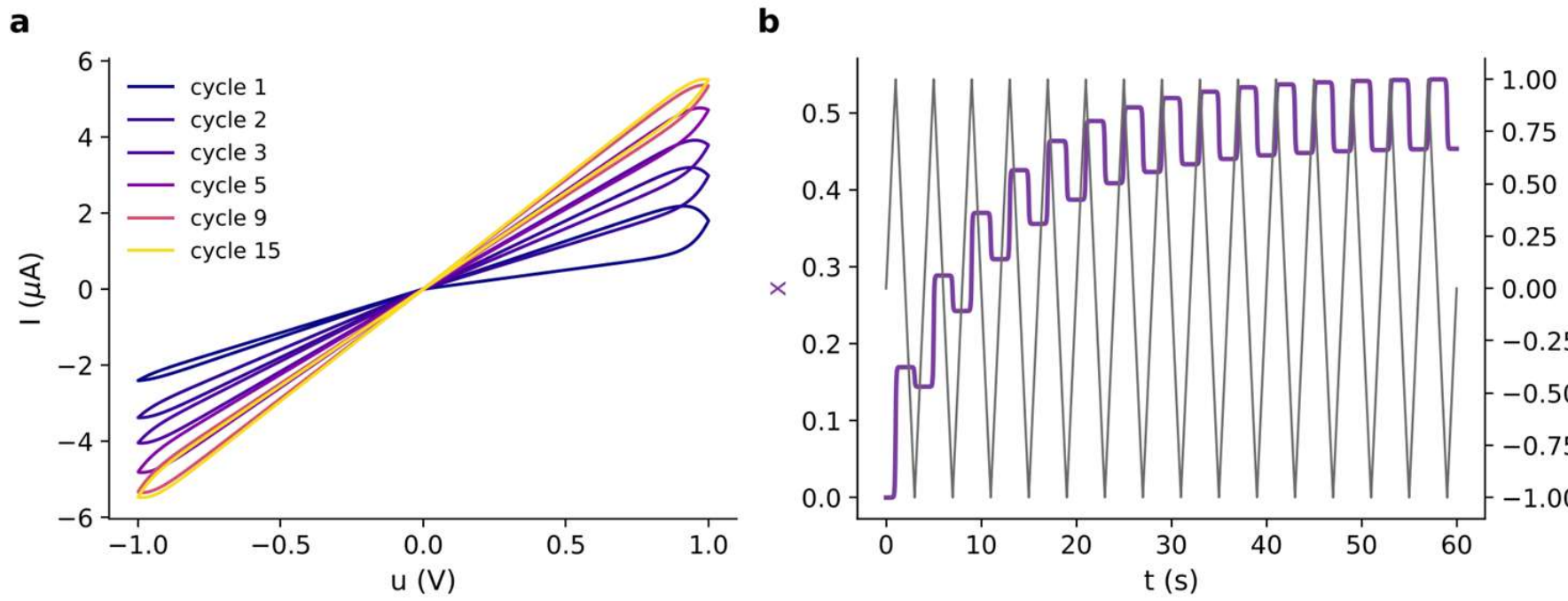

**Figure 5.** Cycling of a gradual bipolar non-volatile device, ±1 V at 1 V/s, starting from $x = 0$. (a) Cycles 1 to 15. (b) State and voltage as a function of time.

The same kinetic law produces regular and inverted hysteresis. For a slow sweep the state lags slightly behind its stationary value, and the current is approximately [15,16]

$$I \simeq I_{ss}(u) + (C_m - g_a\tau)\frac{du}{dt} \tag{6}$$

where $g_a = u\, dG_{ss}/du$ is the relaxing conductance defined in Section 4.5. When an increase of $|u|$ raises the stationary conductance, $g_a > 0$ and the hysteresis is inverted, with the forward current below the reverse current; this is the inductive hysteresis [15]. When it lowers the conductance, $g_a < 0$ and the hysteresis is regular, or capacitive. For a sigmoid centred at 0 V both cases occur in the same loop, at opposite polarities (Fig. 6a). The approximation is verified in Fig. 6b, and the coefficient $-g_a\tau$ is shown in Fig. 6c as a function of bias.

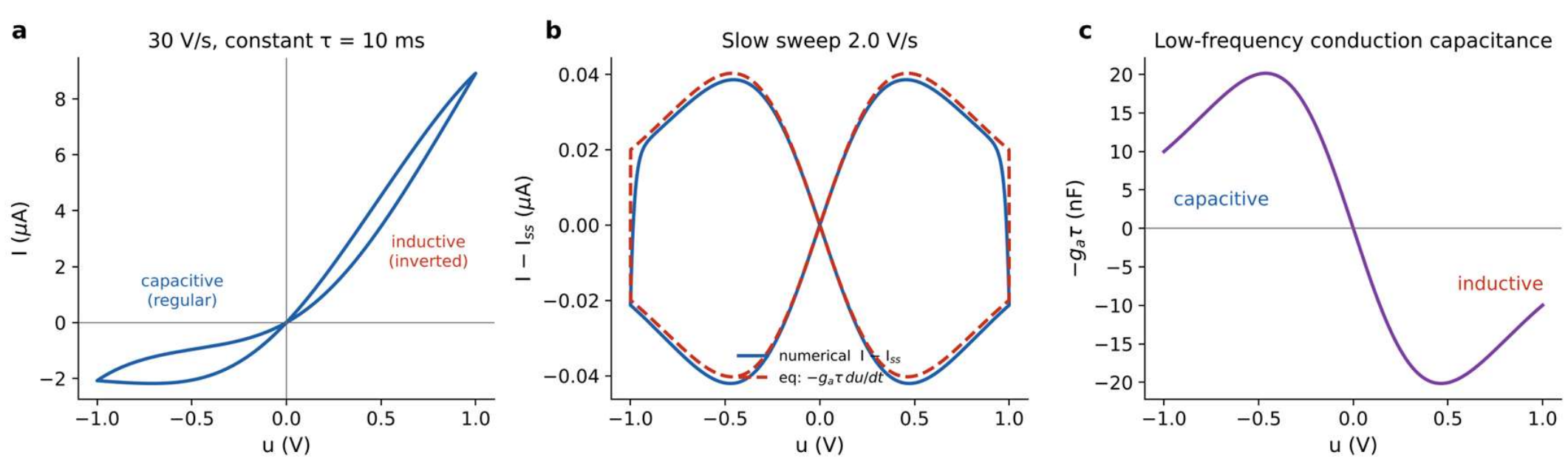

**Figure 6.** Regular and inverted hysteresis from one kinetic law ($x_{eq} = s(u/0.3)$, constant $\tau$ = 10 ms). (a) Loop at 30 V/s. (b) Deviation from the stationary current at 2 V/s compared with $-g_a\tau\, du/dt$. (c) Coefficient $-g_a\tau$, the low-frequency conduction capacitance, as a function of bias.

### 4.4 The CALM model unifies the classes of memristors

With the same equations, and changing only $x_{eq}(u)$ and $\tau(u)$, the model reproduces the main classes of memristors [3]. Figure 7 shows three examples. An abrupt bipolar non-volatile device has a step-like $x_{eq}$ centred at 0 V and a long relaxation time inside a window of ±0.8 V; the state jumps at the window edges (Fig. 7a, 6d, 6g). An abrupt unipolar volatile device has $x_{eq}(|u|)$ with a threshold of 0.8 V and a bell-shaped $\tau$ centred on the switching voltage; since $\tau(0)$ is short, the device returns to HRS at zero bias; volatile and non-volatile responses can even be obtained and separated in the same perovskite device [17] (Fig. 7b, 6e, 6h). A gradual bipolar non-volatile device has a broad $x_{eq}$ and a short-time limit comparable with the sweep, so that intermediate states are produced (Fig. 7c, 6f, 6i). Table 2 summarizes how the class of a device is read from the parameters.

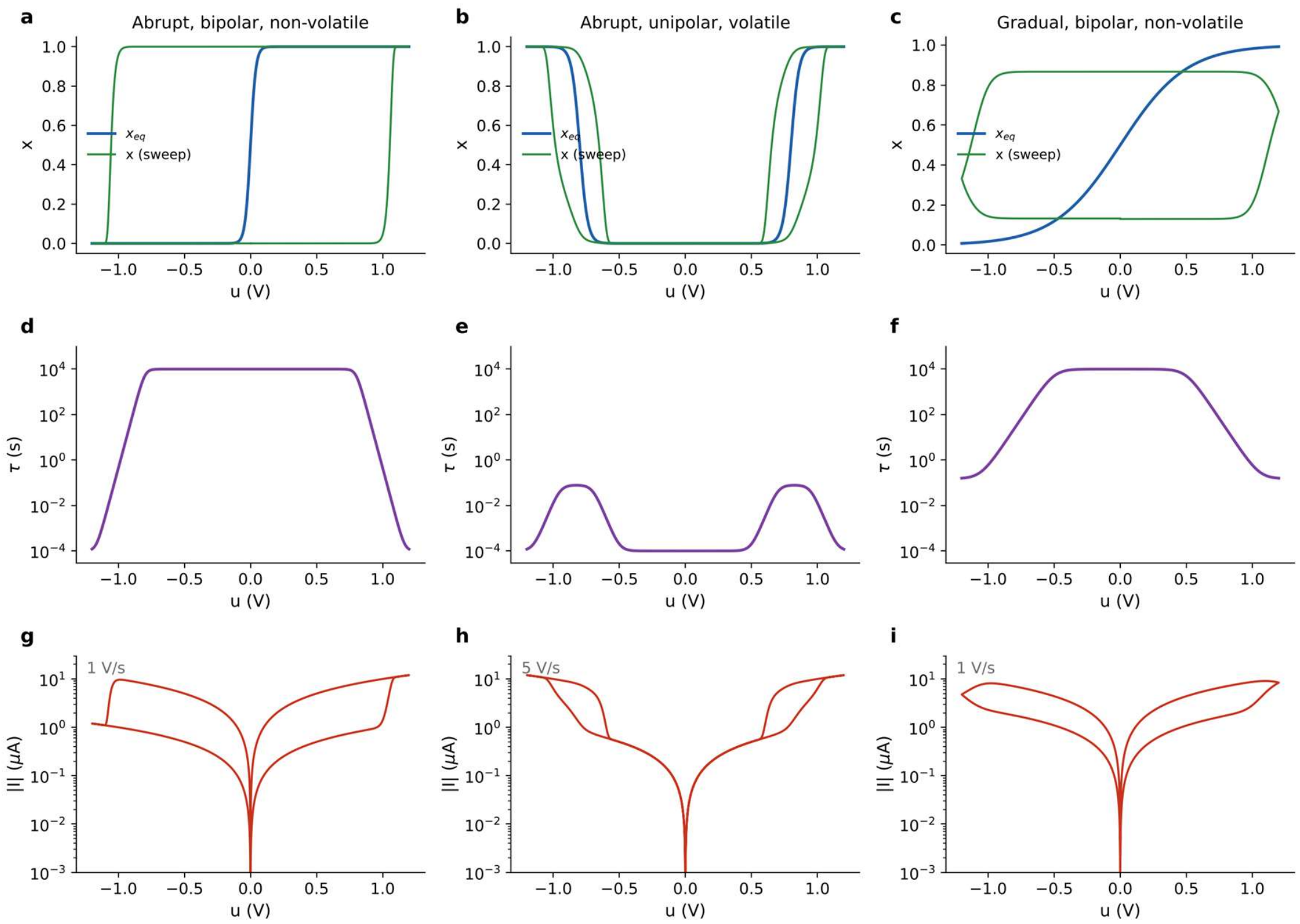

**Figure 7.** Three classes of memristors reproduced by the CALM model. Columns: abrupt bipolar non-volatile (a, d, g), abrupt unipolar volatile (b, e, h) and gradual bipolar non-volatile (c, f, i). Rows: $x_{eq}$ and $x$ along the second sweep cycle (a–c); relaxation time (d–f); logarithm of the absolute current (g–i).

**Table 2. Reading the class of a device from the CALM parameters.** $t_{exp}$ is the time scale of the experiment.

| **Property** | **Controlled by** | **Rule of thumb** |
|---|---|---|
| Bipolar or unipolar | argument of $x_{eq}$ | $x_{eq}(u)$ bipolar; $x_{eq}(|u|)$ unipolar |
| Volatile or non-volatile | $\tau(0)$ | $\tau(0) \ll t_{exp}$ volatile; $\tau(0) \gg t_{exp}$ non-volatile |
| Abrupt or gradual | $V_m$, $\tau_{min}$ | small values abrupt; large values gradual |
| Switching voltages | $V_B$, $V_{ON}$, $V_{OFF}$ | transition where $x_{eq}$ changes and $\tau$ is short |

| Property | Controlled by | Rule of thumb |
|---|---|---|
| On/off ratio | $g_H/g_L$ | read on the logarithmic current plot |
| Hysteresis vs scan rate | $\tau$ near the transition | maximum when the crossing time is about $\tau$ |

### 4.5 Impedance: conduction capacitance and inductance

Impedance spectroscopy applies a small perturbation at a stationary bias $u_0$. Linearizing the CALM equations around $x_0 = x_{eq}(u_0)$ requires three local quantities: the instantaneous conductance $g_b = G(x_0)$, the slope $a = dx_{eq}/du$ at $u_0$ and the time $\tau_0 = \tau(u_0)$ [16]. The relaxing conductance and the differential conductance of the stationary curve are

$$g_a = u_0\,(g_H - g_L)\,a, \qquad G_0 = \frac{dI_{ss}}{du} = g_b + g_a \tag{7}$$

and the admittance is

$$Y(\omega) = g_b + \frac{g_a}{1+i\omega\ {}_0} + i\omega C_m \tag{8}$$

The measured capacitance and the real part of the admittance are

$$C(\omega) = \frac{\operatorname{Im} Y}{\omega} = C_m - \frac{g_a\tau_0}{1+(\omega\tau_0)^2}, \qquad \operatorname{Re} Y(\omega) = g_b + \frac{g_a}{1+(\omega\tau_0)^2} \tag{9}$$

The admittance can be represented by an equivalent circuit [16,18]. For $g_a > 0$ the relaxing branch is a resistance $R_a = 1/g_a$ in series with an inductance $L_a = \tau_0/g_a$. This is the chemical inductor [18]: there is no magnetic element, and the inductance only expresses that the conduction current follows the voltage with a delay. It is therefore called a conduction inductance. For $g_a < 0$ the same branch has a negative inductance, which in the spectrum appears as an additional capacitance, the conduction capacitance [16]; the low-frequency capacitance is then $C_m - g_a\tau_0 > C_m$. Only $\tau_0$ enters the small-signal response; the voltage dependence of $\tau$ does not.

Figure 8 shows the response of a CALM device with $C_m$ = 0.1 nF. At −0.4 V the spectrum has two capacitive arcs (Fig. 8a), with $g_b$ = 2.88 µS and $g_a$ = −1.98 µS; the low-frequency capacitance exceeds $C_m$ by $-g_a\tau$ = 19.8 nF, two hundred times $C_m$. At +0.4 V the low-frequency part is an inductive loop (Fig. 8b). The capacitance spectrum changes sign with the bias, and the conduction contribution vanishes at 0 V (Fig. 8c).

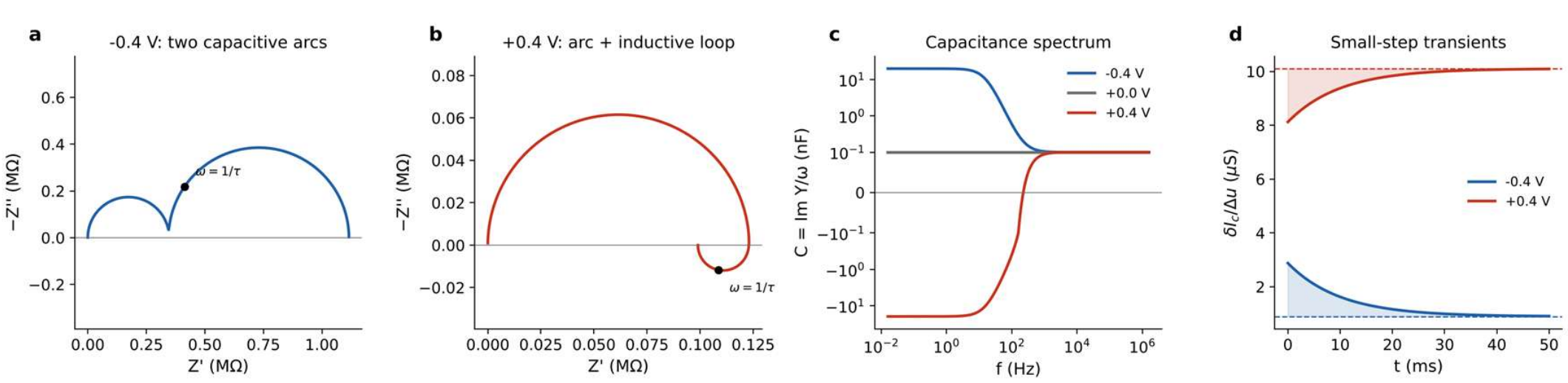

**Figure 8.** Impedance and transients of the CALM model ($x_{eq} = s(u/0.3)$, $\tau$ = 10 ms, $C_m$ = 0.1 nF). (a) Complex-plane plot at −0.4 V. (b) Complex-plane plot at +0.4 V; the dots mark $\omega = 1/\tau$. (c) Capacitance spectra at −0.4, 0 and +0.4 V. (d) Conduction current after

a small voltage step, normalized by the step; dashed lines mark the final values.

The same information is contained in the response to a small voltage step $\Delta u$, for which the conduction current changes as

$$\frac{\delta I_c(t)}{\Delta u} = g_b + g_a\left[1 - e^{-t/\tau_0}\right] \tag{10}$$

Immediately after the step the state has not changed and the conductance is $g_b$; after a time much longer than $\tau_0$ it is $G_0$. When $g_a > 0$ the current rises slowly toward its final value, as in a circuit with an inductor; when $g_a < 0$ it overshoots and decays, as when a capacitor is charged through a resistor (Fig. 8d) [15].

The conduction elements can take values that look surprising if they are interpreted as real components. For a filament or a pore with a conductance of 1 nS and a relaxation time of 1 s, the conduction inductance is of the order of $10^9$ H and the conduction capacitance of the order of 1 nF, much larger than the geometric capacitance of a small device [16]. Observing such large values, and a low-frequency capacitive arc that becomes an inductive loop as the bias increases, is good evidence that the memristive mechanism is being measured [10,16]. A further check is provided by the stationary current–voltage curve, since in the model the amplitude of the relaxing branch is

$$g_a = \frac{dI_{ss}}{du} - \frac{I_{ss}}{u} = u\frac{dG_{ss}}{du} \tag{11}$$

so that the values of $g_a$ obtained from the current–voltage curve, from the impedance and from step transients should agree. The experimental validation of the model with impedance data of halide perovskite and $HfO_2$ memristors is shown in the slides [10,16,19].

### 4.6 Potentiation and depression

In synaptic measurements, voltage pulses emulate presynaptic spikes and the current is the postsynaptic response [20,21]. In the CALM model, a pulse of amplitude $V_p$ and duration $t_p$ followed by an interval $t_i$ at zero bias produces the map [22]

$$x_{n+1} = x_0(1-\beta) + \beta(1-\alpha)\,x_p + \alpha\beta\,x_n \tag{12}$$

with $x_p = x_{eq}(V_p)$, $x_0 = x_{eq}(0)$, $\alpha = e^{-t_p/\tau(V_p)}$ and $\beta = e^{-t_i/\tau(0)}$. The conductance approaches geometrically the fixed point

$$x_F = \frac{x_0(1-\beta)+\beta(1-\alpha)\,x_p}{1-\alpha\beta} \tag{13}$$

Positive pulses produce potentiation and negative pulses depression (Fig. 9a), and the simulation coincides with the map (Fig. 9b). In a volatile device the effect of a pulse decays between pulses, and the paired-pulse facilitation decreases when the interval exceeds $\tau$ (Fig. 9c).

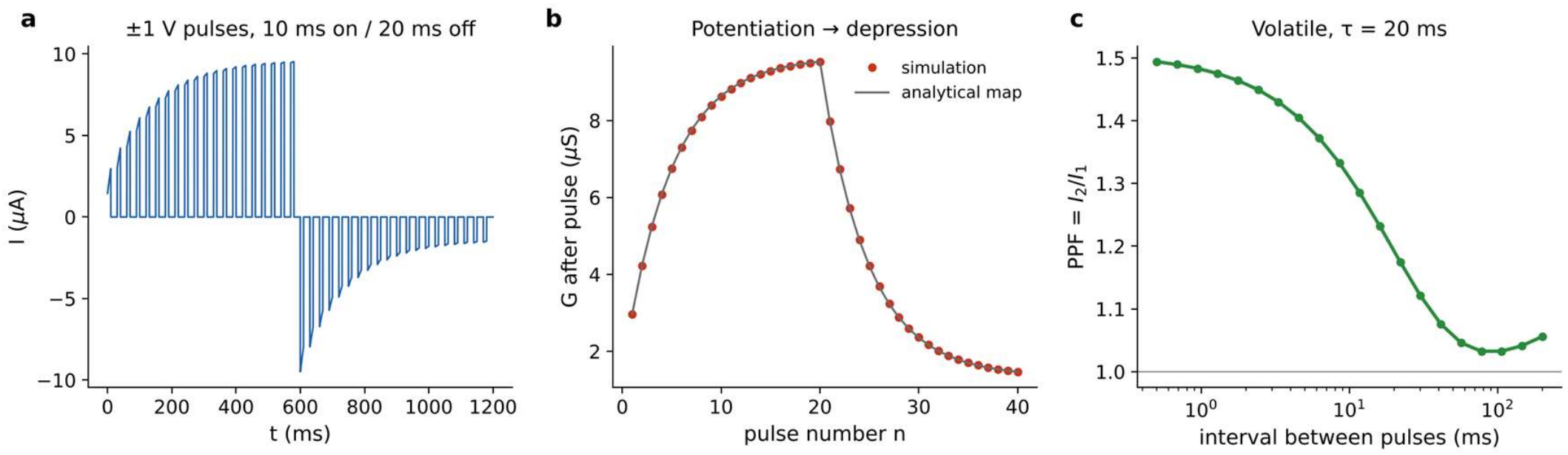

**Figure 9.** Pulse trains in the CALM model. (a) Current under 20 pulses of +1 V followed by 20 pulses of −1 V, 10 ms on and 20 ms off, for a non-volatile device. (b) Conductance at the end of each pulse compared with the analytical map. (c) Paired-pulse facilitation of a volatile device ($x_{eq} = s\big((u-0.5)/0.1\big)$, $\tau$ = 20 ms) as a function of the interval between two 5 ms pulses.

The response to pulse trains also depends on the pulse frequency. In biological synapses, trains at low frequency weaken the connection and trains at high frequency strengthen it, with a threshold frequency between both, as described by the Bienenstock–Cooper–Munro (BCM) rule. This behaviour has been observed in halide perovskite single-crystal memristors [23]. A model with two internal processes reproduces the shape qualitatively if the depressing process switches on fast and recovers slowly, while the potentiating process needs several pulses to build up and recovers fast. At low frequency only the depression accumulates; at high frequency the potentiation also accumulates and dominates (Fig. 10a), and the weight change after the train crosses zero at a threshold frequency (Fig. 10b).

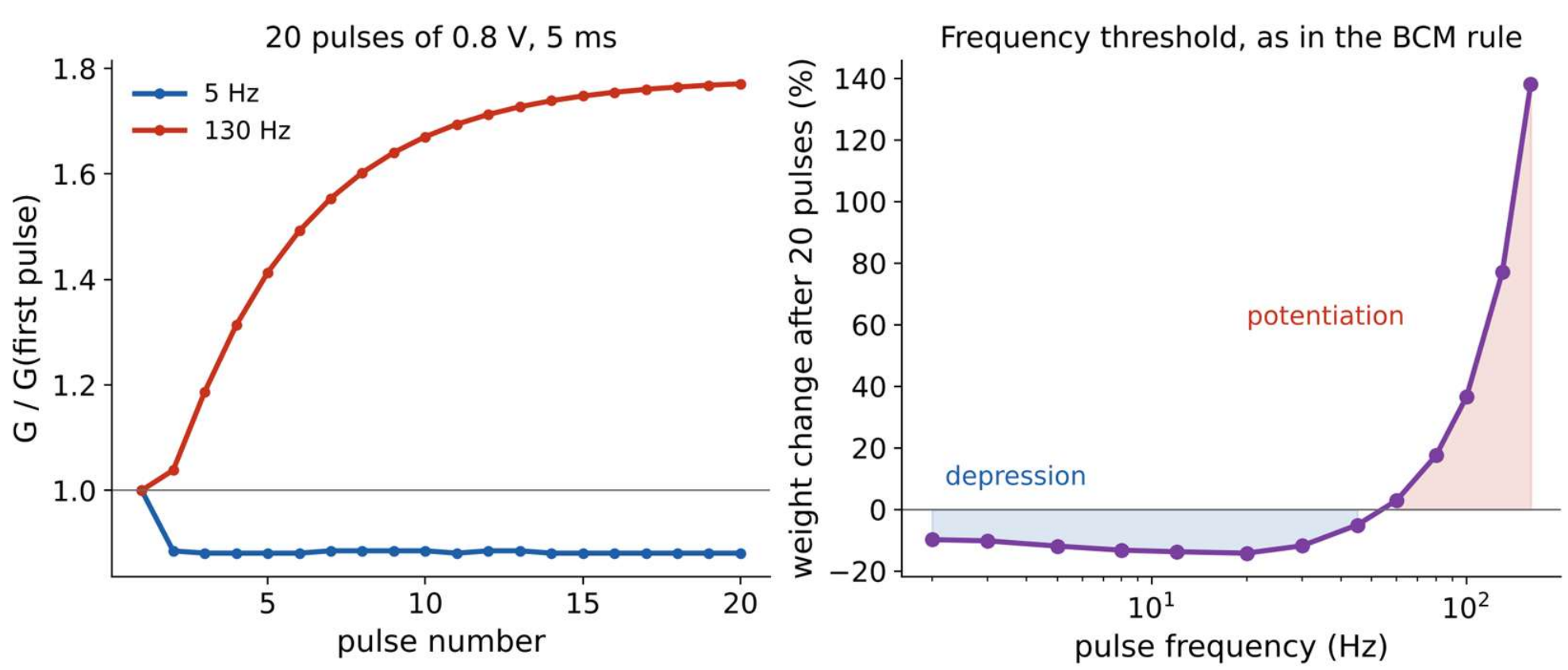

**Figure 10.** Frequency-dependent plasticity in a two-process model ($\tau_d$ = 2 ms during the pulse and 0.5 s at rest; $\tau_p$ = 80 ms during the pulse and 15 ms at rest; 20 pulses of 0.8 V and 5 ms). (a) Conductance during trains at 5 Hz and 130 Hz. (b) Relative change of conductance after the train as a function of the pulse frequency.

The connection with impedance is general [16,20,24]. For any voltage-controlled memristor $I = G(x)u$, $\tau_k\, dx/dt = g(u, x)$, the admittance has the same form, with

$$g_a = \frac{u\,G_x\,g_u}{-g_x}, \qquad \tau = \frac{\tau_k}{-g_x} \tag{14}$$

where subscripts denote partial derivatives. Since $g_a = u\,G_x\,dx_{ss}/du$, it is positive when an increase of the voltage magnitude raises the stationary conductance, the small-signal counterpart of potentiation, and negative under depression. Potentiation is therefore accompanied by a rising current after a step, a negative capacitance, an inductive loop and inverted hysteresis; depression by an overshoot, a positive capacitance and regular hysteresis [15,16,24].

### 4.7 Beyond the simplest picture

**Non-pinched loops.** The statement that a memristor shows a hysteresis loop pinched at the origin [25] is a theorem about the ideal equation $I = G(x,u)\,u$, which gives $I = 0$ whenever $u = 0$. Real devices carry other currents in parallel: the displacement current of the geometric capacitance, the charging current of ionic accumulation at interfaces, and diffusion currents driven by ion concentration gradients. With a parallel capacitance the loop opens at $u = 0$ by $C_m\,du/dt$, which grows linearly with the scan rate (Fig. 11b, 9d). With an ionic branch made of a resistance $R_s$ in series with a capacitance $C_s$, a residual current remains at zero bias and is maximum when the sweep time is comparable with $R_s C_s$ (Fig. 11c, 9d). The ideal model is pinched at every rate (Fig. 11a). Non-zero crossing loops of this kind have been observed in perovskite memristors with an ion-modulating interfacial layer at high compliance currents and fast sweeps [26]. A non-pinched loop does not disqualify a device as a memristor; it indicates that some element stores charge, and the model should include it.

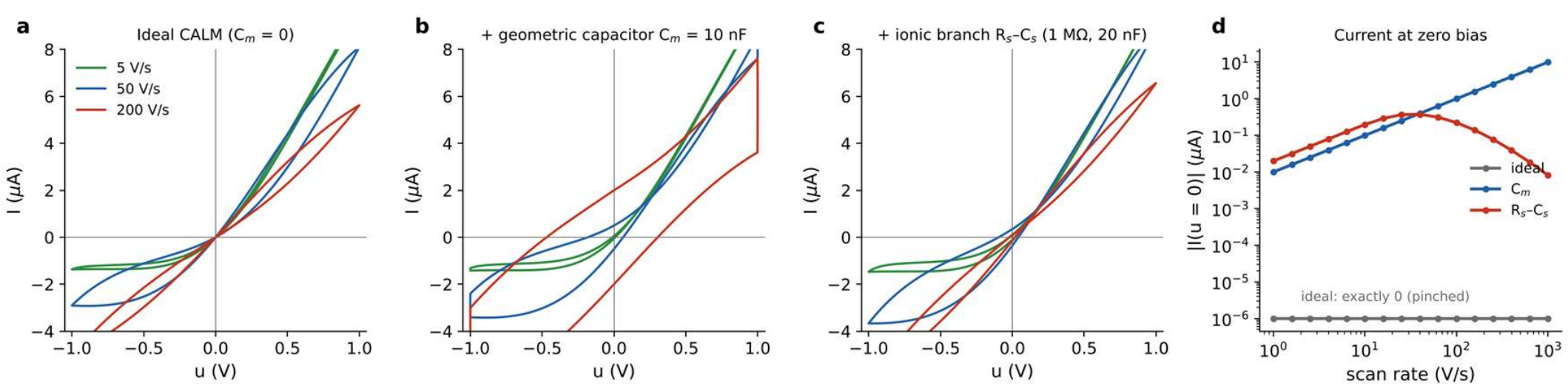

**Figure 11.** Opening of the loop by an additional capacitor ($x_{eq} = s(u/0.3)$, $\tau$ = 10 ms). (a) Ideal CALM model. (b) With a parallel capacitance $C_m$ = 10 nF. (c) With an ionic branch $R_s$ = 1 MΩ, $C_s$ = 20 nF. (d) Absolute current at $u = 0$ as a function of the scan rate.

**Several internal variables.** Real devices often contain two or more processes with different thresholds and times. The model generalizes directly,

$$I = G(x_1, \dots, x_n)\,u + C_m \frac{du}{dt}, \qquad \tau_i(u)\frac{dx_i}{dt} = x_{i,eq}(u) - x_i \tag{15}$$

and each variable adds one relaxing branch to the admittance,

$$Y(\omega) = g_b + \sum_i \frac{g_{a,i}}{1+i\omega\tau_i} + i\omega C_m, \qquad g_{a,i} = u_0\,\frac{\partial G}{\partial x_i}\frac{dx_{i,eq}}{du} \tag{16}$$

The relaxing conductances can have opposite signs, so that capacitive and inductive features coexist and the sign of the hysteresis can change within one branch of the current–voltage curve.

**Crossing at 0.5 V.** As an example, consider a unipolar device with a depressing variable $x_d$, threshold 0.25 V, and a potentiating variable $x_p$, threshold 0.88 V,

$$G = g_0 - \Delta g_d\, x_d + \Delta g_p\, x_p \tag{17}$$

with $\Delta g_d$ = 1.5 µS, $\Delta g_p$ = 10 µS, $g_0$ = 2.5 µS and $\tau$ = 20 ms for both variables. The threshold of $x_p$ was chosen so that the stationary conductance has its minimum at 0.5 V; the relaxing conductance is negative below 0.5 V and positive above (Fig. 12a). The hysteresis is regular below 0.5 V and inverted above, and the forward and reverse branches cross at 0.50 V at 1 V/s, 0.49 V at 2 V/s and 0.46 V at 4 V/s (Fig. 12b). Trains of pulses of 0.3 V produce depression and trains of 0.8 V produce potentiation (Fig. 12c), and the impedance changes from a capacitive arc to an inductive arc (Fig. 12d). A transition from capacitive to inductive behaviour as the voltage increases produces a non-zero crossing of the current–voltage curves [16]. In general the slow-sweep crossing occurs where $\sum_i g_{a,i}\,\tau_i = 0$, so that it coincides with the minimum of $G_{ss}$ only when the times are equal.

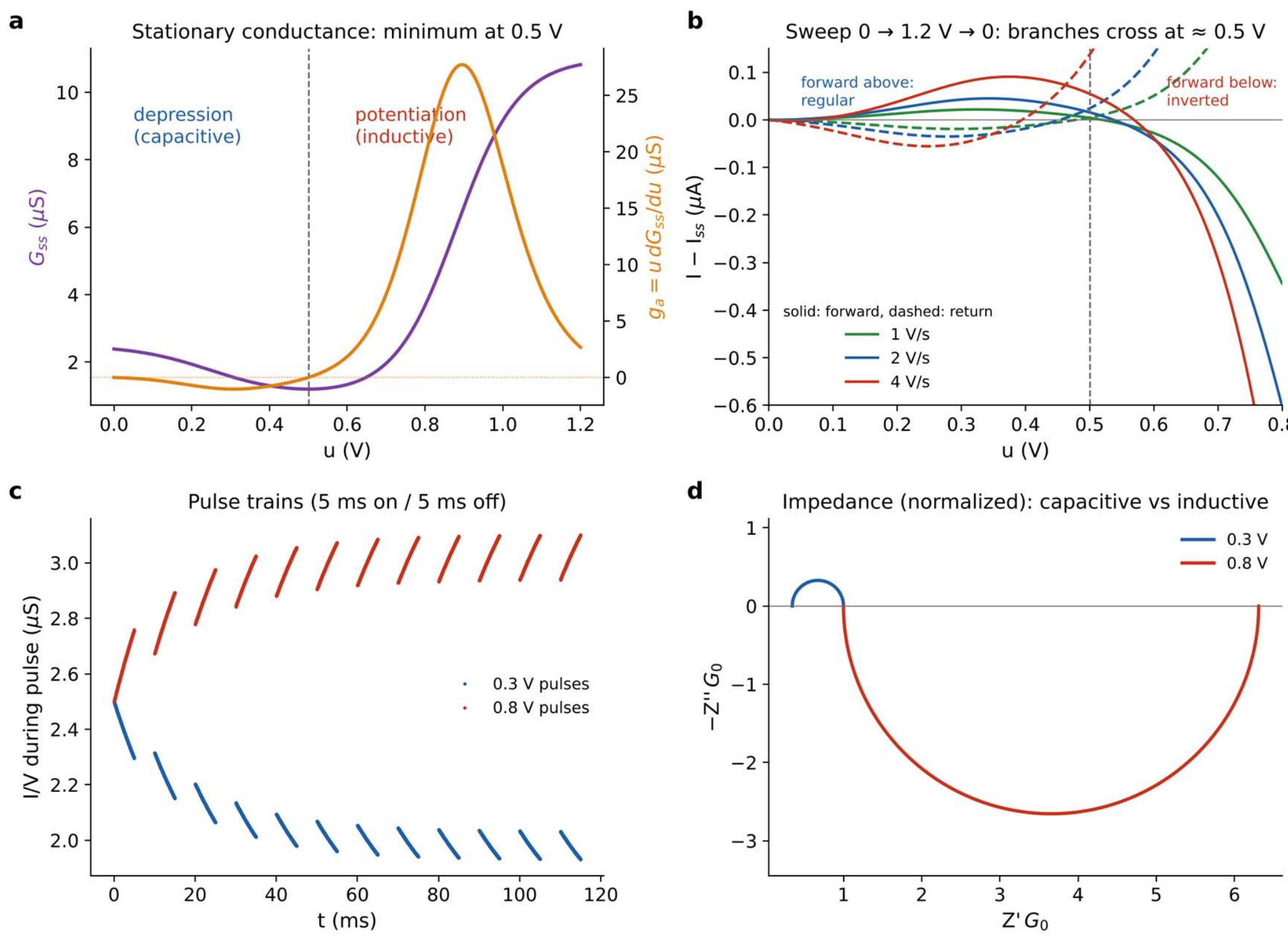

**Figure 12.** Two internal variables with opposite effects. (a) Stationary conductance and total relaxing conductance. (b) Deviation of the current from $I_{ss}$ for sweeps 0 → 1.2 V → 0 at three rates; solid lines forward, dashed lines return. (c) Conductance during trains of 5 ms pulses at 0.3 V and 0.8 V. (d) Normalized impedance at 0.3 V and 0.8 V.

**Dual-inductive response.** If the state responds to $|u|$ and raises the conductance, the product $u\, G_x\, g_u$ is positive at both polarities. The loops are inverted in both quadrants and the impedance shows the same inductive loop at $\pm u_0$ (Fig. 13a, 11b). This unipolar volatile CALM device describes the dual-inductive switching observed in halide perovskite memristors with ion-modulated interfacial barriers [26].

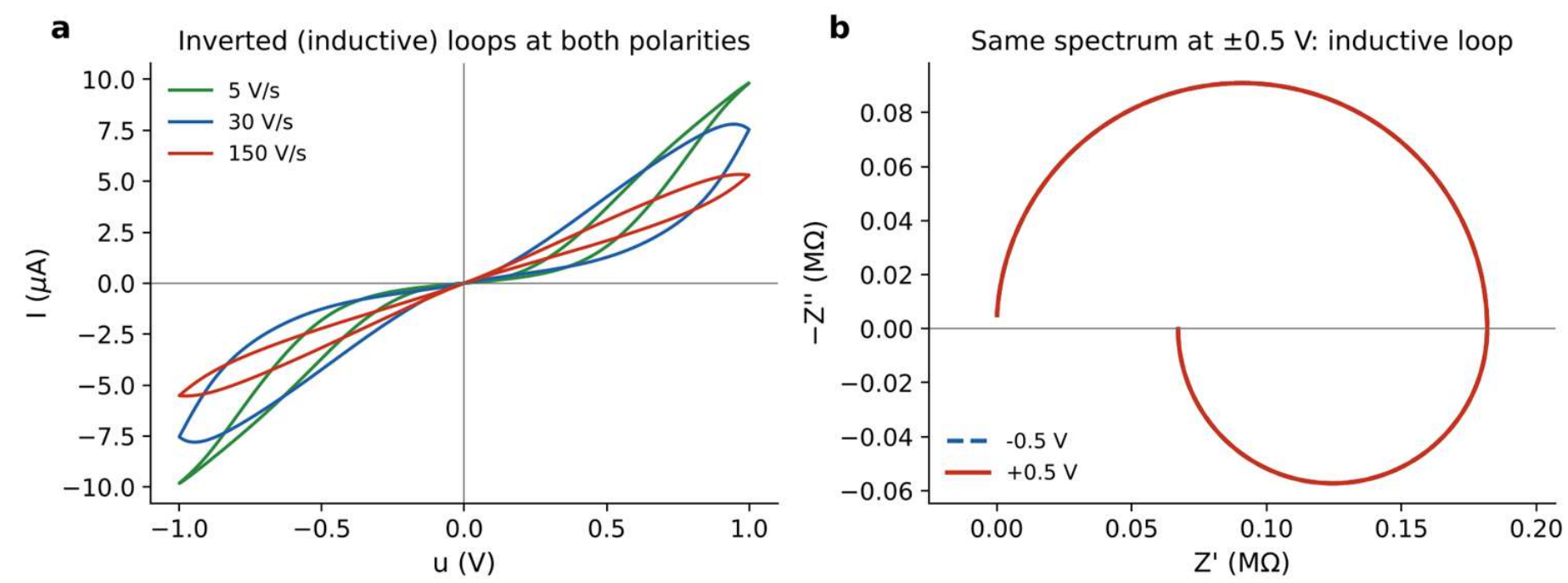

**Figure 13.** Unipolar volatile CALM device. (a) Inverted loops at both polarities. (b) Complex-plane impedance plots at −0.5 V and +0.5 V, which coincide.

### 4.8 Applying the CALM model to experimental data

A practical workflow has six steps. (1) Measure the stationary current–voltage curve by slow sweeps or held voltages, read $g_L$ and $g_H$ from the two linear limits, and fit $x_{eq}(u)$ to obtain $V_B$ and $V_m$. (2) Measure loops at three or four decades of scan rate; the position of the maximum hysteresis gives $\tau$ near the transition, and the full set constrains $\tau(u)$. (3) Hold the device at 0 V and read the decay of the conductance to obtain $\tau(0)$ and decide whether the device is volatile. (4) Measure impedance spectra or small-step transients at several biases to obtain $g_b$, $g_a$ and $\tau_0$. (5) Cross-check: $g_a$ from impedance must agree with $u\, dG_{ss}/du$ from step 1, its sign must follow the bias, and the sign of the hysteresis must follow $C_m - g_a\tau$. (6) Predict the response to pulse trains with the map of Section 4.6 and compare with measurements.

Common pitfalls are fitting a single loop, which cannot separate $x_{eq}$ from $\tau$; ignoring the cycle history of non-volatile devices; identifying every capacitive arc with a stored charge and every inductive loop with an artefact; forcing a pinched model onto a non-pinched loop; and describing with one variable a loop that crosses away from zero, which usually requires two processes with relaxing conductances of opposite sign.

## 5. Key equations

Table 3 collects the equations used in the tutorial.

**Table 3. Key equations of the CALM model.**

| Quantity | Equation |
|---|---|
| Current | $I = [g_L + (g_H - g_L)x]u + C_m\, du/dt$ |
| State | $\tau(u)\, dx/dt = x_{eq}(u) - x$ |

| Quantity | Equation |
|---|---|
| Rectification function | $x_{eq} = 1/\{1 + \exp[-(u - V_B)/V_m]\}$ |
| Stationary current | $I_{ss} = G(x_{eq})\, u$ |
| Relaxing conductance | $g_a = u\, dG_{ss}/du$ |
| Admittance | $Y = g_b + g_a/(1 + i\omega\tau_0) + i\omega C_m$ |
| Low-frequency conduction capacitance | $C(0) - C_m = -g_a\tau_0$ |
| Slow-sweep hysteresis | $I \simeq I_{ss} + (C_m - g_a\tau)\, du/dt$ |
| Pulse map | $x_{n+1} = x_0(1-\beta) + \beta(1-\alpha)x_p + \alpha\beta x_n$ |

## Acknowledgments

This work was funded by the European Research Council (ERC) via Horizon Europe Advanced Grant, grant agreement nº 101097688 ("PeroSpiker").

## References

(1) Chua, L. O.; Kang, S. M. Memristive Devices and Systems. *Proc. IEEE* **1976**, 64, 209–223.

(2) Bisquert, J. Hysteresis, Rectification, and Relaxation Times of Nanofluidic Pores for Neuromorphic Circuit Applications. *Adv. Phys. Res.* **2024**, 3, 2400029.

(3) Bou, A.; Gonzales, C.; Boix, P. P.; Vaynzof, Y.; Guerrero, A.; Bisquert, J. Kinetics of Volatile and Nonvolatile Halide Perovskite Devices: The Conductance-Activated Quasi-Linear Memristor (CALM) Model. *J. Phys. Chem. Lett.* **2025**, 16, 69–76.

(4) Kim, S.-Y.; Rivera-Sierra, G.; Mengesha, B. S.; Iniguez, B.; Bisquert, J. Lead-Free Bismuth Halide Perovskite Memristors: Low-Voltage Switching and Physical Modeling of Resistive Hysteresis. *Adv. Mater. Technol.* **2026**, e02152.

(5) Rivera-Sierra, G.; Ramirez, P.; Bisquert, J.; Bou, A. Relaxation Time of Multipore Nanofluidic Memristors for Neuromorphic Applications. *J. Am. Chem. Soc.* **2025**, 147, 17529–17538.

(6) Solanki, A.; Guerrero, A.; Zhang, Q.; Bisquert, J.; Sum, T. C. *J. Phys. Chem. Lett.* **2020**, 11, 463–470.

(7) Kim, S.-Y.; et al. *J. Phys. Chem. Lett.* **2024**, 15, 10087–10103.

(8) Lee, Y.; Park, J.; Chung, D.; et al. Multi-Level Cells and Quantized Conductance Characteristics of $Al_2O_3$-Based RRAM Device for Neuromorphic System. *Nanoscale Res. Lett.* **2022**, 17, 84.

(9) Ramirez, P.; Gómez, V.; Cervera, J.; Mafe, S.; Bisquert, J. Synaptical Tunability of Multipore Nanofluidic Memristors. *J. Phys. Chem. Lett.* **2023**, 14, 10930–10934.

(10) Gonzales, C.; Bou, A.; Guerrero, A.; Bisquert, J. Capacitive and Inductive Characteristics of Volatile Perovskite Resistive Switching Devices with Analog Memory. *J. Phys. Chem. Lett.* **2024**, 15, 6496–6503.

(11) Xiao, Y.; et al. *Adv. Devices Instrum.* **2024**, 5.
(12) Hodgkin, A. L.; Huxley, A. F. A Quantitative Description of Membrane Current and Its Application to Conduction and Excitation in Nerve. *J. Physiol.* **1952**, 117, 500–544.
(13) Cole, K. S.; Baker, R. F. Longitudinal Impedance of the Squid Giant Axon. *J. Gen. Physiol.* **1941**, 24, 771–788.
(14) Chua, L.; Sbitnev, V.; Kim, H. Hodgkin–Huxley Axon Is Made of Memristors. *Int. J. Bifurcation Chaos* **2012**, 22, 1230011.
(15) Bisquert, J. Inductive and Capacitive Hysteresis of Current-Voltage Curves: Unified Structural Dynamics in Solar Energy Devices, Memristors, Ionic Transistors, and Bioelectronics. *PRX Energy* **2024**, 3, 011001.
(16) Bisquert, J.; Roldán, J. B.; Miranda, E. Hysteresis in Memristors Produces Conduction Inductance and Conduction Capacitance Effects. *Phys. Chem. Chem. Phys.* **2024**, 26, 13804–13813.
(17) Pendyala, N.-K.; Gonzales, C.; Guerrero, A. Decoupling Volatile and Nonvolatile Response in Reliable Halide Perovskite Memristors. *Small Struct.* **2024**, 2400380.
(18) Bisquert, J.; Guerrero, A. Chemical Inductor. *J. Am. Chem. Soc.* **2022**, 144, 5996–6009.
(19) Marquardt, R.; et al. *Adv. Electron. Mater.* **2023**, 2201227.
(20) Bisquert, J.; Shim, W.; Kim, S.-Y.; Linares-Barranco, B. Synaptic Function in Memristor Devices for Neuromorphic Circuit Applications. *Adv. Electron. Mater.* **2025**, 11, 2400903.
(21) Zhou, L.; et al. *Cell Rep. Phys. Sci.* **2024**, 5, 102078.
(22) Rivera-Sierra, G.; Bisquert, J. From Pulses to Plasticity: Analytical Tools for Memristive Synapse Design. *APL Mach. Learn.* **2025**, 3, 046102.
(23) Matchenya, I.; Khanas, A.; Podgornyi, R.; Shirkin, D.; Ekgardt, A.; Sizykh, N.; Anoshkin, S.; Krasnikov, D. V.; Yulin, A.; Zhukov, A.; Nasibulin, A. G.; Scheblykin, I. G.; Pushkarev, A.; Zenkevich, A.; Bisquert, J.; Marunchenko, A. Short-Term Bienenstock-Cooper-Munro Learning in Optoelectrically-Driven Flexible Halide Perovskite Single Crystal Memristors. *Small Methods* **2025**, 2500203.
(24) Kim, S.-Y.; Zhang, H.; Rivera-Sierra, G.; Fenollosa, R.; Rubio-Magnieto, J.; Bisquert, J. Introduction to Neuromorphic Functions of Memristors: The Inductive Nature of Synapse Potentiation. *J. Appl. Phys.* **2025**, 137, 111101.
(25) Chua, L. O. If It's Pinched It's a Memristor. *Semicond. Sci. Technol.* **2014**, 29, 104001.
(26) Kim, S.-Y.; Bisquert, J. Dual-Inductive and Programmable Switching: A New Paradigm in Ionic Interface-Controlled Perovskite Memory. *Adv. Funct. Mater.* **2026**, e75888.

## Solutions to the exercises

The numerical values below were obtained with *calm_tutorial.py*. Results from other codes should agree within the accuracy of the integration.

### S1. Build the stationary curve

The stationary current is not a straight line: it follows $g_L u$ below the threshold and $g_H u$ above it (Fig. 1a), so the device rectifies even without dynamics. With $V_m$ = 0.02 V the transition is a kink, with $V_m$ = 0.25 V a smooth curve (Fig. 1b). For $V_m$ = 0.08 V, $I(1\text{ V})$ = 9.98 µA and $I(-1\text{ V})$ = −1.00 µA, so the ratio of currents is $g_H/g_L$ = 10. For $V_m$ = 0.25 V the ratio is lower, 8.93/1.02, because $x_{eq}(1\text{ V})$ = 0.88 has not reached 1.

### S2. Your first hysteresis loop

At 0.5 V/s the loop collapses onto $I_{ss}$ (Fig. 2a). At 20 and 200 V/s the state lags behind $x_{eq}$ (Fig. 2b): during the forward sweep $x < x_{eq}$ and the current is lower than in the reverse sweep, so the hysteresis at positive voltage is inverted (inductive). The loop passes through the origin because $I = G(x)\,u$ vanishes at $u = 0$ whatever the value of $x$; with $C_m = 0$ the loop is pinched. At 200 V/s a small lobe appears at negative voltage, because $x$ is still high when the voltage crosses zero.

### S3. Scan rate and the time constant

The loop area as a function of the logarithm of the scan rate is a bell-shaped curve with a maximum at about 150 V/s (Fig. 4b). Multiplying $\tau_{max}$ by 10 shifts the maximum by one decade, to about 15 V/s, with almost the same height: the position of the maximum scales as $1/\tau$. The loop disappears at slow rates because $x$ follows $x_{eq}$, and at fast rates because $x$ does not move.

### S4. Cycling and the sign of the loop

(a) The state at the end of each cycle increases as 0.14, 0.24, 0.31, 0.36, … and converges to 0.455; it is within 1% of the limit after about 13 cycles (Fig. 5). (b) The numerical deviation $I - I_{ss}$ and the approximation $-g_a\tau\, du/dt$ coincide (Fig. 6b), which shows that the slow-sweep hysteresis is controlled by the single coefficient $g_a\tau$. The hysteresis is regular in the negative quadrant ($g_a < 0$) and inverted in the positive quadrant ($g_a > 0$).

### S5. Reproduce the three classes

One set of parameters is given in Fig. 7 and in *calm_tutorial.py* (parameter sets B1, B2, B3). (i) is converted into a volatile device by making $\tau(0)$ short, for instance by moving the window away from zero; the stored state then decays at zero bias. In (ii), a constant $\tau$ equal to the bell maximum makes the state respond slowly also away from the threshold, and the loop broadens (its area at 5 V/s grows from 4.2 to 5.2 µW); a constant $\tau$ equal to the minimum removes the hysteresis almost completely. The logarithmic plots reproduce the shapes of the volatile, non-volatile gradual and non-volatile abrupt devices of ref [3].

**S6. Impedance from the model**

At −0.4 V: $g_b$ = 2.877 µS, $g_a$ = −1.981 µS, $G_0$ = 0.896 µS and $-g_a\tau$ = +19.8 nF; the spectrum shows a high-frequency arc due to $C_m$ and a second, larger capacitive arc (Fig. 8a). At +0.4 V: $g_b$ = 8.123 µS, $g_a$ = +1.981 µS, $G_0$ = 10.10 µS and $-g_a\tau$ = −19.8 nF; the low-frequency part is an inductive loop (Fig. 8b). At 0 V, $g_a = 0$ and only the arc of $C_m$ in parallel with $1/g_b$ remains. The low-frequency plateaus of $C(\omega)$ are $C_m \pm 19.8$ nF (Fig. 8c).

**S7. Pulses, potentiation and facilitation**

(a) With $\alpha = e^{-0.01/0.05}$ = 0.819, $\beta = e^{-0.02/2}$ = 0.990, $x_p$ = 0.982 and $x_0 = x_{eq}(0)$ = 0.5, the fixed point is $x_F$ = 0.957, and the conductance at the end of a pulse saturates at about 9.65 µS (Fig. 9b). It remains below $g_H$ because during each interval the state relaxes toward $x_{eq}(0)$ = 0.5; a longer off-time lowers $x_F$. Under −1 V pulses the conductance decreases to about 1.4 µS. (b) With $x_{eq} = s\big((u - 0.5)/0.1\big)$, so that the device rests in HRS, the paired-pulse ratio is about 1.5 for short intervals and decays to about 1 when the interval exceeds a few times $\tau$ = 20 ms (Fig. 9c), so $\tau$ can be read from the decay. More analog levels are obtained with shorter pulses, smaller $\alpha$ steps, or a smaller pulse amplitude that brings $x_p$ closer to the middle of the sigmoid.

**S8. Break the pinch**

(i) $I(0) = 0$ at all rates (Fig. 11a). (ii) $I(0) = C_m\, du/dt$: 0.1 µA at 10 V/s and 1.0 µA at 100 V/s, growing linearly with the rate (Fig. 11b, 9d). (iii) The residual current at zero bias has a maximum of about 0.38 µA near 30 V/s (Fig. 11c, 9d). At this rate the time to sweep from the vertex to 0 V, about 30 ms, is comparable with $R_s C_s$ = 20 ms: at slower rates the capacitor follows the voltage, and at faster rates it has no time to charge. Experimentally, a $C_m$ effect grows indefinitely with the rate and is independent of bias history, while an ionic branch has a characteristic rate and usually depends on the preconditioning and the compliance current [26].

**S9. Design a crossing**

The condition $dG_{ss}/du = 0$ at 0.5 V gives $V_P$ = 0.881 V. The stationary conductance is 2.39 µS at 0 V, 1.57 µS at 0.3 V, 1.20 µS at 0.5 V and 3.67 µS at 0.8 V (Fig. 12a). The branches cross at 0.50, 0.49 and 0.46 V for 1, 2 and 4 V/s (Fig. 12b). Pulses at 0.3 V depress and pulses at 0.8 V potentiate (Fig. 12c). With $\tau_p = 5\tau_d$ the slow-sweep crossing moves to the voltage where $g_{a,d}\tau_d + g_{a,p}\tau_p = 0$, about 0.42 V; the simulation gives 0.40 V at 0.5 V/s and 0.33 V at 1 V/s, and at faster sweeps the crossing disappears because the slow potentiating variable dominates the whole branch.

## Collaborators

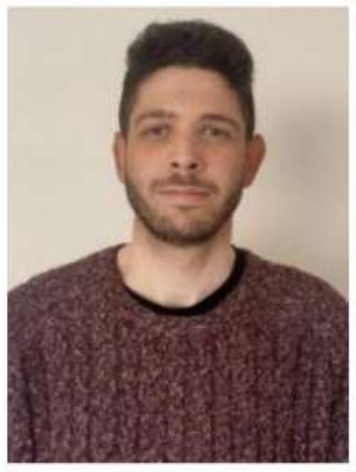

**Agustín Bou**

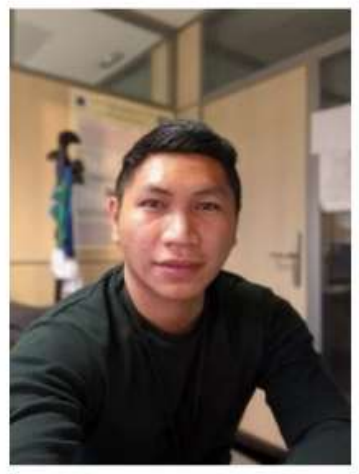

**Cedric Gonzales**

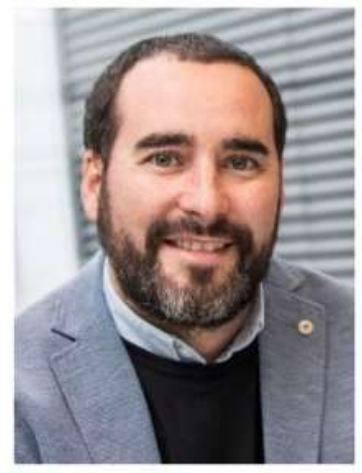

**Antonio Guerrero**

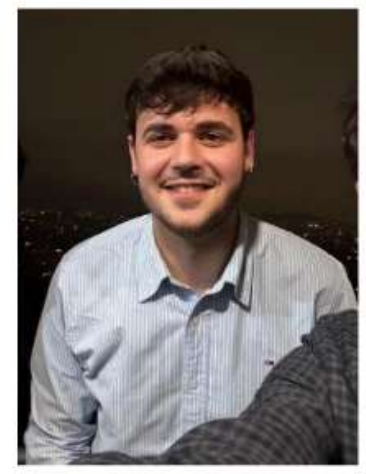

**Gonzalo Rivera-Sierra**

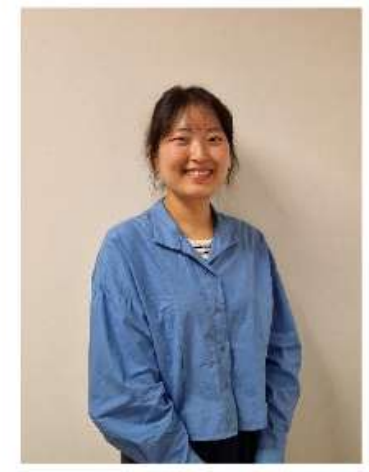

**So-Yeon Kim**

Instituto de Tecnología Química (UPV-CSIC), València, and collaborating groups. The models, experiments and many figures of this tutorial come from our joint work.